\documentclass[twocolumn,aps,amssymb,showpacs,nofootinbib]{revtex4}
\usepackage{graphicx}
\usepackage{float}
\usepackage[usenames,dvipsnames]{color}
\usepackage{natbib}
\usepackage{multirow}
\usepackage{ulem}
\usepackage{setspace}

 \fontsize{15}{15}

\begin{document}
\title{False-positive and False-negative assignments of Topological Insulators in Density-Functional Theory and Hybrids}
\date{\today}
\author{J. Vidal}
\author{X. Zhang}
\author{L. Yu}
\author{J.-W. Luo}
\author{A. Zunger}
\email{alex.zunger@gmail.com}
\affiliation{National Renewable Energy Laboratory, Golden, Colorado 80401, USA}

\begin{abstract}
Density-functional Theory (DFT) approaches have recently been used to judge the topological order of various materials despite its well-known band gap underestimation. Use of the more accurate quasi-particle GW approach reveals here few cases where DFT identifications are false-positive, possibly misguiding experimental searches of materials that are topological insulators (TI) in DFT but not expected to be TI in reality. We also present the case of false-positive due to the incorrect choice of crystal structures and adress the relevancy of such choice of crystal structure with respect to the ground state one and thermodynamical instability with respect to binary competing phases. We conclude that it is then necessary to consider both the correct ground state crystal structure and the correct Hamiltonian in order to predict new TI.
\end{abstract}


\maketitle
3D bulk topological insulators (TI) have a band inversion, e.g. the $s$-like conduction band below the ($p,d$)-like valence band at the time reversal invariant momentum such as $\Gamma$. This order of bands is defined through the inversion energy $\Delta_i=\varepsilon_s-\varepsilon_{p,d}$ being negative (see Figure 1). Specifically, for normal insulators such as CdTe, $\Delta_i$ is positive while for topological insulators such as HgTe, $\Delta_i$ is negative. The excitation gap defined as the energy difference between the highest occupied state and the lowest unoccupied state can be either zero (Fig 1b) or positive (Fig 1c) even in TI. When one tuncates a 3D bulk TI to create a 2D surface or interface, new states will appear inside the 2D excitation band gap: these will have linear dispersion (`massless fermions`) crossing each other ('gapless') within the 2D excitation gap and maintain spin-polarization without net charge current.

Interest in their exotic properties has motivated recent search and discovery efforts of materials that would be topological insulators~\cite{chadov2010tunable,lin2010half,PhysRevLett.105.096404,PhysRevB.82.235121,PhysRevB.82.125208,PhysRevLett.105.216406, PhysRevLett.106.156402,PhysRevLett.106.016402}. Theory predicts interesting cases expected to be TI's and pleas are then made to experimentalists to make and probe such materials in the laboratory. As exciting as such a close level of theory-experiment interaction is, there are two potential issues that can impede progress. First, most of these calculations rely on density functional theory (DFT) or variants of it~\cite{chadov2010tunable,lin2010half,PhysRevLett.105.096404,PhysRevB.82.235121,PhysRevB.82.125208,PhysRevLett.105.216406, PhysRevLett.106.156402,PhysRevLett.106.016402}, yet these methods are known to systematically underestimate band gaps specifically by placing the conduction band too low and the valence band too high, thereby creating the potential of 'false positive' prediction of a material being TI in DFT but not in reality, to the detriment of experiment-theory interaction. Second, sometimes~\cite{chadov2010tunable,lin2010half,PhysRevLett.105.096404,PhysRevB.82.235121,PhysRevB.82.125208,PhysRevLett.105.216406, PhysRevLett.106.156402,PhysRevLett.106.016402} theoretical predictions of negative inversion energy $\Delta_i$ are conducted on \textit{assumed} crystal structures of given candidate materials; in some cases those assumed structures are energetically quite far from the ground state structure. This results in predicting an exciting property (e.g., TI) in unrealizable structures: we present here a case where the assumed crystal structure makes the compound thermodynamically unstable with respect to dissociation in competing binary phases.   

In this Communication, we revisit the topological order of several families of materials previously predicted to be TI within DFT,  but now we are using (i) improved approximations for the electronic levels GW and (ii) insist on calculating topological order in structures that are ground states or very close to it. The group of materials we study are: (a) binary compounds InAs and HgTe (b) Half Heusler compounds ABX [A=(Y,Lu); B=(Pt,Pd) and X=(Sb,Bi)]~\cite{chadov2010tunable,lin2010half}, (c) Antiperovskite nitride (M)$_3$BiN [(M=Ca,Sr,Ba)]~\cite{PhysRevLett.105.216406}, and (d) Honeycomb-lattice chalcogenides LiAgSe and NaAgSe~\cite{PhysRevLett.106.156402}. We first describe the method we used based on quasiparticle self-consistent GW. Then, we present a list of false-positive cases due to DFT and as false-positives due to the wrong crystal structure. Finally, we assess the potential of the recently used~\cite{chen2011band} screened Hybrid functional, HSE06~\cite{HSE06} with respect to both DFT and GW methods. 

\textit{Methods:} Contrary to DFT, quasi-particle (QP) energies can be compared directly with experiments. Also, GW does not suffer from self-interaction error and is orbital dependent\cite{onida2002electronic}, which are essential for excitation gap prediction. Out of the many flavors of GW available, we employ the fully self-consistent (sc-) version of GW\cite{PhysRevLett.96.226402}. The need for self-consistence is obvious in the case of topological insulator: usually GW is employed within the so-called G$_0$W$_0$ approximation, which is a perturbative correction to energies obtained from a mean-field Hamiltonian (usually DFT). However, such approximation assumes implicitly that the mean-field Hamiltonian could describe the system of interest accurately enough so that its energies are considered to be good approximation to quasiparticle energies~\cite{bruneval2006effect}. DFT dramatic underestimation of band gaps could result in metallic band structure within DFT for narrow band gap semiconductors. Therefore, G$_0$W$_0$ is not effective in such cases and one need to self-consistently update both QP energies and wavefunctions in the GW Hamiltonian. Furthermore, spin orbit splittings calculated using norm-conserving relativistic pseudopotentials~\cite{PhysRevB.58.3641} were added to the GW quasiparticle energies as a first order perturbation. All GW calculations were performed using the crystal structure relaxed within generalized gradient approximation (GGA).

\textit{False-positive predictions of topological characteristics due to DFT}: Calculated DFT and sc-GW inversion energies for previously proposed TI materials are given in Table I, together with their topological designation.

\textit{(a) InAs and HgTe}: InAs is a narrow-excitation-gap normal semiconductor with $\Delta^{\rm Expt.}_i$=\ 0.43 eV. Sc-GW reproduces it well (see Table I) while DFT predicts a negative inversion energy and consequently identify it incorrectly as TI.

 Therefore, the assignement of topological order to InAs within DFT is considered a \textit{false positive} and is due to the strong underestimation of the band gap in DFT. In the case of HgTe, both $\Delta_i^{\rm sc-GW}$ and $\Delta_i^{\rm DFT}$  
are negative, making HgTe a topological insulator in agreement with experimental findings. Only sc-GW is able to qualitatively and quantitatively reproduce experimental inversion energy and excitation gap for both HgTe and InAs.

(b) \textit{The Filled Tetrahedral Structures (FTS)}: ABX with A=(Y,Lu); B=(Pd,Pt) and X=(Sb,Bi) crystallize in a structure based on the zinc blende lattice which has a fourfold coordinated cation at position T$_1$=(0,0,0) and a fourfold coordinated anion at T$_2$=($\frac{1}{4}$,$\frac{1}{4}$,$\frac{1}{4}$). Whereas in the zinc blende lattice,
the two tetrahedral intersitial sites - IT$_1$ at $\left(\frac{1}{2},\frac{1}{2},\frac{1}{2}\right)$ 
coordinated by four T$_2$ anions, and IT$_2$ at $\left(\frac{3}{4},\frac{3}{4},\frac{3}{4}\right)$ 
coordinated by four T$_1$ cations - are empty, the FTS are based 
on stuffing an additional atom either at IT$_1$ (called $\alpha$-type FTS~\cite{PhysRevB.31.2570,PhysRevLett.56.528}) 
or at IT$_2$ ($\beta$-type FTS~\cite{PhysRevB.31.2570,PhysRevLett.56.528}). We study both types. We found twice more false positive in $\alpha$-type than in $\beta$-type. Such difference can be attributed to the localization of the VBM wavefunction: in $\alpha$-type, VBM is formed by non -bonding states, highly localized around atoms while in $\beta$-type, VBM is formed by more delocalized bonding states. As a result, $\alpha$-type FTS suffers more from DFT-induced self-interaction error~\cite{PhysRevB.23.5048}: VBM is predicted to lie too high in energy within DFT, closing the band gap and in some extreme cases inverting the order of the bands around Fermi level and therefore incorrectly assign it to be TI.

\textit{(c) Strained antiperovskite nitrides M$_3$BiN (M=Ca,Sr and Ba)}:
The materials belonging to this family are said to be topological insulators under strain (but not as strain free). According to DFT calculations, unstrained M$_3$BiN (M=Ca, Sr and Ba) antiperovskites have a small energy gap between low-lying unoccupied $s$-states of Bi and occupied $p$-states of Bi (defined as the inversion energy $\Delta_i$), prompting the suggestion~\cite{PhysRevLett.105.216406} that application of external strain can drive such small gap systems to have TI-like negative inversion energy. Band gaps obtained within sc-GW for unstrained bulk are in good agreement with experimental data when available~\cite{chern1992synthesis}. Such GW calculation for M$_3$BiN (M=Ca,Sr,Ba) showed that the energy separation between $p$-like valence states and $s$-like conduction states significantly increases from $\Delta^{\rm DFT}_i=0.65$ eV to $\Delta^{\rm sc-GW}_i=1.85$ eV for Ca$_3$BiN, from $\Delta^{\rm DFT}_i=0.26$ eV to $\Delta^{\rm sc-GW}_i=1.55$ eV for Sr$_3$BiN and from $\Delta^{\rm DFT}_i=1.08$ eV to $\Delta^{\rm sc-GW}_i=1.79$ eV for Ba$_3$BiN, making gap closure by strain extremely unlikely. Therefore, strained (Ca,Sr,Ba)$_3$BiN are probably \textit{false positive}.

\textit{False Positive due to the wrong choice of crystal structure}: LiAgSe and NaAgSe in the assumed honeycomb-lattice (ZrBeSi-type structure  see Fig 2a) are one of the few DFT-predicted TI materials with light elements ($Z<50$). The VBM is formed by $p$-$d$ hybridized states from Se and Ag while the conduction band minimum (CBM) is exclusively $s$-like. Within DFT, LiAgSe in the ZrBeSi-structure has a negative inversion energy at $\Gamma$ while NaAgSe in the same structure presents an odd number of negative inversion energy at $\Gamma$, A and L, which makes both of them TIs~\cite{PhysRevLett.106.156402} within such crystal structure and DFT. Within sc-GW, none of the negative inversion energy are observed, leading to the conclusion that NaAgSe and LiAgSe are normal insulators  in the ZrBeSi-structure (see table I for values of inversion energy at $\Gamma$). This is similar to the false positive due to DFT presented in the previous section. 

Unfortunately, ZrBeSi-type structure assumed for (Li,Na)AgSe in reference~\cite{PhysRevLett.106.156402} is not the ground state structure. According to our total energy calculations performed on more than 50 potential crystal structures, LiAgSe crystallizes in a LiCaN-structure (Figure 2b) (0.085 eV/atom lower than ZrBeSi-type) and NaAgSe crystallizes in the PbCl$_2$-type structure (Figure 2c) (0.05 eV/atom lower than ZrBeSi-type). Moreover, LiAgSe in the ZrBeSi-type structure  is unstable with respect to dissociation into Li$_2$Se and Ag$_2$Se.

Interestingly, both DFT and sc-GW predict a positive inversion energy for (Li,Na)AgSe in their respective ground state structure at any \textbf{k}-point in the Brillouin zone. Therefore, we discover a very special case of \textit{false positive} due to the incorrect choice of ground-state structure, independently of the type of methods used to predict the inversion energies.

\begin{figure}
\begin{center}
 \includegraphics[scale=0.45]{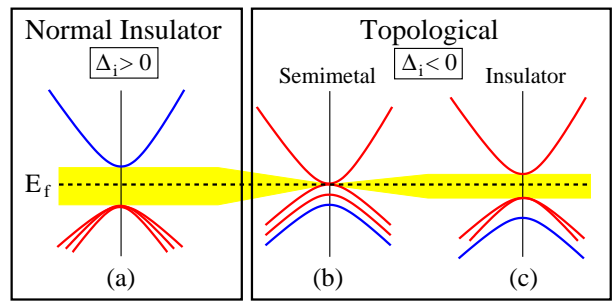}
\caption{(color online). Schematic of the band ordering at the vicinity of Fermi level for (a) normal insulator, (b) topological semimetal and (c) topological insulator. Red color represents (p,d)-like character while blue color represents s-like character.}
\end{center}
\end{figure}

\begin{figure}
\begin{center}
 \includegraphics[scale=0.12]{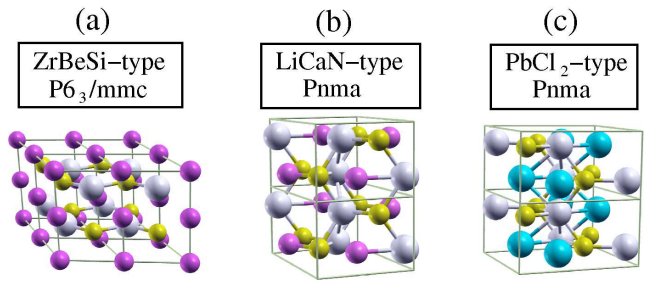}
\caption{(color online). Structure type for (a) ZrBeSi-type (b) LiCaN-type and (c) PbCl$_2$-type.}
\end{center}
\end{figure}

\textit{Is HSE06 the answer ?}

Recently, Chen \textit{et al.}~\cite{chen2011band} determine new topological insulators based on screened hybrid functional HSE06~\cite{HSE06}, which is known to improve band gaps of solids with respect to DFT~\cite{paier2006screened}. We performed HSE06 calculations for all the materials in Table I and CuTl(S,Se)$_2$~\cite{note5} recently predicted to be TI~\cite{PhysRevLett.106.016402}. For most compounds, HSE06 gives qualitatively and quantitatively very similar results compared to sc-GW, with a 20$\%$ deviation for $\Delta_i$ with respect to sc-GW results. However, two problematic cases appear where HSE06 is in qualitative disagreement with both DFT and sc-GW: HgTe where $\Delta^{\rm HSE06}_i \approx 0$ and the determination of topological order is then uncertain and CuTlS$_2$ where $\Delta^{\rm HSE06}_i > 0$ (see Table II) yet $\Delta^{sc-GW}<0$. Such \textit{false negative} cases occurs because of the dependence of HSE06 on a fixed mixing parameter for exchange, $a_x= \frac {1}{4}$. The mixing parameter $a_x$ could be linked to inverse dielectric constant in the long range limit $a_x \approx \frac{1}{\epsilon_\infty}$~\cite{maksimov1989excitation,PhysRevLett.104.056401,PhysRevB.83.035119}. In the case of HSE06 and HgTe, $a_x = \frac{1}{4}$ and $\epsilon^{\rm HgTe}_\infty \approx 20.0$.

In conclusion, we revisited some of the numerous theoretical predictions of TI and found that many of them are subject to \textit{false positive} and \textit{false negative} cases due to an inadequate use of \textit{ab initio} methods or assumed structure. We use a robust method based on GW in order to avoid false positive and false negative. We adress the importance of using both the ground state structure and the correct Hamiltonian in order to search for new TIs. The failure to do so could result in false prediction and might mislead experimentalists in their effort to grow new TIs.  

We thank Dr S. Lany for fruitful discussions. Research supported by the U.S. Department of Energy,
Office of Basic Sciences, Division of Materials Sciences
and Engineering, under
Grant No. DE-AC36-08GO28308 to NREL.

 \begin{table*}[h!]
\caption{Calculated DFT and $GW$ inversion energies $\Delta_i$, and topological class (T=topological and N=normal) for InAs, HgTe, ($\alpha$-,$\beta$-)(Y,Lu)(Pd,Pt)(Sb,Bi), (Li,Na)AgSe in the assumed ZrBeSi-type structure and their ground state structure.}
\begin{center}
 \begin{tabular}{lcccc}
\hline
\hline
\multirow{3}{*}{Material (Structure)} & \multicolumn{2}{c}{Inversion energy } & \multicolumn{2}{c}{\multirow{2}{*}{Topological ?}} \\
 & \multicolumn{2}{c}{[eV]} &  \\
 & DFT & sc-GW & DFT & sc-GW  \\
\hline
InAs (zb)& -0.6 & 0.65 & T & N \\
HgTe (zb)& -1.1 & -0.26 & T & T \\
\hline
LuPtSb ($\beta-$FTS)\footnote[1]{calculated ground state} & -0.21 & 1.04 & T & N \\
YPdBi ($\beta$-FTS)\footnote[1]{calculated ground state} & -0.09 & 0.5 & T & N  \\
LuPtSb ($\alpha$-FTS) & -0.65 & 0.45 & T & N  \\
LuPdBi ($\alpha$-FTS)& -0.27 & 0.52 & T & N  \\
YPtSb ($\alpha$-FTS) & -0.5 &  0.45 & T & N  \\
YPdBi ($\alpha$-FTS) & -0.1 & 0.45 & T & N \\
\hline
LiAgSe (Fig. 2a)\footnote[2]{Excited state structure 0.085 eV/atom higher} & -0.17 & 1.3 & T &  N  \\
LiAgSe (Fig. 2b)\footnote[3]{Ground state structure} & 0.4 & 1.9 & N &  N  \\

NaAgSe (Fig. 2a) \footnote[4]{Excited state structure 0.05 eV/atom higher} & -0.23  & 0.8 & T &  N  \\
NaAgSe (Fig. 2c) \footnote[5]{Ground state structure} & $\sim$0.0 & 0.75 & N &  N  \\
\hline
\hline
\end{tabular}
\end{center}
\end{table*}

 \begin{table*}[h!]
\caption{Calculated DFT, HSE06 and $GW$ inversion energies $\Delta_i$, and topological class (T=topological and N=normal) for CuTl(S,Se)$_2$. CH stands for chalcopyrite.}
\begin{center}
 \begin{tabular}{ccccccc}
\hline
\hline
\multirow{2}{*}{Material (Structure)}  &  \multicolumn{3}{c}{Inversion energy [eV]} &  \multicolumn{3}{c}{Topological ?} \\

  & DFT & HSE06 & sc-GW  & DFT & HSE06 & sc-GW \\
\hline
CuTlS$_2$ (CH) & -0.77 &  0.06 & -0.41 & T & N & T  \\
CuTlSe$_2$ (CH) & -1.0 & -0.4  & -0.85 & T & T & T \\
\hline
\hline
\end{tabular}
\end{center}
\end{table*}


\begin{thebibliography}{24}
\bibitem[{not()}]{note5}
\bibinfo{note}{Both CuTl(S,Se)$_2$ were found to be TI within sc-GW. However,
  in order to obtain a qualitative estimation of the inversion energy, we used
  HSE06+G$_0$W$_0$ approach which is known to be equivalent to
  sc-GW~\cite{PhysRevB.76.115109} and allows us to treat metallicity of the
  band structure.}
\end{thebibliography}
\end{document}